\documentclass{ws-procs9x6-cpt16} 
\begin{document} 

\newcommand*\al{\alpha}
\newcommand*\be{\beta}
\newcommand*\ga{\gamma}
\newcommand*\de{\delta} 
\newcommand*\ep{\epsilon}
\newcommand*\ve{\varepsilon}
\newcommand*\ze{\zeta}
\newcommand*\et{\eta}
\newcommand*\vt{\vartheta}
\newcommand*\io{\iota}
\newcommand*\ka{\kappa}
\newcommand*\la{\lambda} 
\newcommand*\vpi{\varpi}
\newcommand*\rh{\rho}
\newcommand*\vrho{\varrho}
\newcommand*\si{\sigma}
\newcommand*\vs{\varsigma}
\newcommand*\ta{\tau}
\newcommand*\up{\upsilon}
\newcommand*\ph{\phi}
\newcommand*\vp{\varphi}
\newcommand*\ch{\chi}
\newcommand*\ps{\psi}
\newcommand*\om{\omega}
\newcommand*\Ga{\Gamma}
\newcommand*\De{\Delta}
\newcommand*\Th{\Theta}
\newcommand*\La{\Lambda}
\newcommand*\Si{\Sigma}
\newcommand*\Up{\Upsilon}
\newcommand*\Ph{\Phi}
\newcommand*\Ps{\Psi}
\newcommand*\Om{\Omega}
\newcommand*\mn{{\mu\nu}} 

\newcommand*\pt[1]{\phantom{#1}}
\newcommand*\prt{\partial}
\newcommand*\vev[1]{\langle {#1}\rangle}
\newcommand*\expect[1]{\langle{#1}\rangle}
\newcommand*\bra[1]{\langle{#1}|}
\newcommand*\ket[1]{|{#1}\rangle}
\newcommand*\fr[2]{{\textstyle{{#1} \over {#2}}}}
\newcommand*\half{{\textstyle{1\over 2}}}
\newcommand*\lsim{\mathrel{\rlap{\lower4pt\hbox{\hskip1pt$\sim$}}
    \raise1pt\hbox{$<$}}}
\newcommand*\gsim{\mathrel{\rlap{\lower4pt\hbox{\hskip1pt$\sim$}}
    \raise1pt\hbox{$>$}}}
\newcommand*\sqr[2]{{\vcenter{\vbox{\hrule height.#2pt
         \hbox{\vrule width.#2pt height#1pt \kern#1pt
         \vrule width.#2pt}
         \hrule height.#2pt}}}}
\newcommand*\lrprt{\stackrel{\leftrightarrow}{\partial}}
\newcommand*\lrprtmu{\stackrel{\leftrightarrow}{\partial_\mu}}
\newcommand*\lrprtnu{\stackrel{\leftrightarrow}{\partial^\nu}}
\newcommand*\lrD{\stackrel{\leftrightarrow}{D}}
\newcommand*\lrDmu{\stackrel{\leftrightarrow}{D_\mu}}
\newcommand*\lrDnu{\stackrel{\leftrightarrow}{D^\nu}}
\newcommand*\lrdla{\stackrel{\leftrightarrow}{D^\la}}

\newcommand*\abs[1]{\left|{#1}\right|}

\newcommand*\tb{\tilde{b}} 
\newcommand*\td{\tilde{d}} 
\newcommand*\tg{\tilde{g}} 
\newcommand*\tc{\tilde{c}} 
\newcommand*\tH{\tilde{H}}

\newcommand{\tablecite}[1]{[\refcite{#1}]} 

\newcommand{\openone}{\pmb{1}} 

\newcommand{\refeq}[1]{(\ref{#1})}
\def\etal {{\it et al.}} 

\title{Spacetime Variation of Lorentz-Violation Coefficents\\
at Nonrelativistic Scale} 

\author{Charles D.\ Lane,$^{1,2}$} 

\address{$^1$Department of Physics, Berry College,\\
Mount Berry, GA, 30149-5004, USA}

\address{$^2$IU Center for Spacetime Symmetries, Indiana University,\\
Bloomington, IN 47405-7105, USA}

\begin{abstract} 
When the Standard-Model Extension (SME) is applied in curved spacetime, 
the Lorentz-violation coefficients must depend on spacetime position. 
This work describes some of the consequences of this spacetime variation. 
We focus on effects that appear at a nonrelativistic scale 
and extract sensitivity of completed experiments 
to derivatives of SME coefficient fields. 
\end{abstract}

\bodymatter

\section{Introduction} 
Within the SME in Minkowski spacetime, 
Lorentz violation is controlled by a set of tensor coefficients 
that do not vary with spacetime position.\cite{sme} 
In curved spacetime, however, 
Lorentz-violation coefficients that vary with spacetime position 
must be considered.\cite{akgravity} 
In this work, 
we describe some of the consequences of the variation of these coefficients, 
particularly at a nonrelativistic scale.\cite{lane2016} 

\section{Nonrelativistic Hamiltonian} 
Our framework is the minimal SME in weakly-curved spacetime 
with nonzero $a_\mu$ and $b_\mu$ coefficient fields, 
which is described by an action 
that is a combination of conventional, Lorentz-symmetric terms 
and a set of terms $\de S$ that includes the effects of nonzero and spacetime-varying 
SME coefficients $a_\mu$ and $b_\mu$: 
\begin{equation} 
\label{action} 
\de S = \int d^4 x \left\{ 
 -\overline{\ps} 
 \left( \de^\mu_a -\half h^\mu_a +\half h^\al_\al \de^\mu_a \right) 
 \left( a_\mu\ga^a + b_\mu\ga_5\ga^a \right) 
 \ps 
 \right\} \quad . 
\end{equation} 

To extract nonrelativistic effects, 
we apply the following procedure: 
(1) Perform a field redefinition to ensure a hermitian hamiltonian, 
 and therefore unitary time evolution. 
(2) Apply the Euler-Lagrange equations to the resulting action, 
 then solve for $i\prt_0\ps=H\ps$ 
 to extract the relativistic $4\times 4$ hamiltonian $H$. 
(3) Perform a Foldy-Wouthuysen transformation, 
 resulting in a nonrelativistic $4\times 4$ hamiltonian $H_{\rm NR}$. 

The full result of these calculations is too long to include in this work, 
but typical behavior appears in the term 
\begin{displaymath}  
H_{\text{NR}} \supset 
 + \pmb{1} \left\{ \left[ 
 -m h^{m0}
 -\frac{i\prt_{[j} b_{k]}}{4m} {\ve_p}^{mj} 
  \left(\de^{kp} - \hbar^{kp}\right) 
 \right] \frac{p_m}{m} 
 \right\} \quad . 
\label{exampleterm}
\end{displaymath} 
This term acts on nonrelativistic matter 
like certain terms in the SME nonrelativistic hamiltonian 
for Minkowski spacetime: 
\begin{displaymath}  
H_{\text{NR,Mink}} 
 \supset \pmb{1} \left\{ \left[mc^{0m} +mc^{m0}\right]\frac{p_m}{m} \right\} 
 \quad . 
\end{displaymath}  
We may exploit this similarity to interpret the weakly-curved-spacetime factors 
as effective versions of the Minkowski-spacetime coefficients: 
\begin{displaymath}  
\left( mc^{0m}+mc^{m0} \right)_{\text{eff}} = 
 -m h^{m0}
 -\frac{i\prt_{[j} b_{k]}}{4m} {\ve_p}^{mj} \left( 
  \de^{kp} - \hbar^{kp} 
  \right) 
\end{displaymath}  
We can then use the results of studies of the Minkowski-spacetime SME 
to interpret the effects of spacetime-varying coefficients. 
When the full calculation is done, 
we find that all terms that result from spacetime-varying $a_\mu$ and $b_\mu$ 
may be interpreted as effective values of 
Minkowski-spacetime coefficients $a_\mu$, $b_\mu$, $c_{\mn}$, and $d_{\mn}$. 

Many derivatives $\prt_\mu a_\nu$ and $\prt_\mu b_\nu$ contribute 
to multiple effective coefficients. 
In Table \ref{dominant}, 
the largest contribution of each derivative to the nonrelativistic hamiltonian 
is displayed. 
\begin{table} 
\tbl{Dominant appearance of each derivative. 
Note that $\prt_0 a_0$, $\prt_0 b_0$, and $\prt_0 b_k$ are absent, 
while only the trace part of $\prt_{(j}a_{k)}$ appears.} 
{
\begin{tabular}{c@{\hspace{0.5cm}}c@{\hspace{0.5cm}}c} 
\text{Minkowski-spacetime} & \text{Weakly-curved-spacetime} & \text{Intuitive} \\ 
\text{coefficient} 
  & \text{coefficient} 
  & \text{equivalent} \\ \hline 
$mc^{0m}+mc^{m0}$ 
  & $-\fr{\prt_{[j} b_{k]}}{4m} {\ve^{mj}}_p \de^{kp}$ 
  & $\text{curl}(\vec{b})$ \\ 
$-mc_{00}$ 
  & $-\fr{i\prt_{(j} a_{k)}}{2m} \de^{jk}$ 
  & $i\, \text{div}(\vec{a})$ \\ 
$md_{q0}$ 
  & $\fr{\prt_{[j} a_{k]}}{2m} {\ve^{jk}}_q$ 
   $+\fr{i\prt q b_0}{2m}$ 
  & $\text{curl}(\vec{a}) +i\, \text{grad}(b_0)$ \\ 
$m\de^{mk}d_{qk}+m\de^m_q d_{00}$ 
  & $\fr{1}{4m}{\ve^{jm}}_q \left( \prt_j a_0 -\prt_0 a_j \right)$ 
  & $\text{grad}(a_0) -\prt_0\vec{a}$ \\ 
 & $+\fr{i\prt_{(j}b_{k)}}{4m}{\ve^{pj}}_q {\ve^{km}}_p$ 
  & $i\prt_{(j} b_{k)}$ \\ 
 &  $-\fr{i\prt_{[j}b_{k]}}{4m} \left( {\ve^{pm}}_q {\ve^{jk}}_p +{\ve^{pj}}_q {\ve^{mk}}_p \right)$ 
  & $i\, \text{curl}(\vec{b})$
\end{tabular} 
} 
\label{dominant} 
\end{table} 

\section{Analysis} 
\subsection{Hermiticity} 
The Foldy-Wouthuysen transformation is unitary, 
and therefore a hermitian relativistic hamiltonian $H$ 
is guaranteed to yield a hermitian nonrelativistic hamiltonian $H_{\rm NR}$. 
However, many individual terms are nonhermitian, 
including terms that are hermitian in the Minkowski spacetime limit. 

The key mathematical idea is that the product $AB$ of hermitian operators $A$ and $B$ 
is hermitian if and only if $A$ and $B$ commute with each other. 
For example, 
fermion momentum operators $p_j$ commute with 
the SME coefficient $b_0$ 
in Minkowski spacetime 
since $b_0$ does not vary with spacetime position. 
However, 
in curved spacetime, 
$\prt_j b_0 \ne 0$, 
and therefore $[p_j,b_0]\ne 0$. 
As a result, the term 
$-\ga^j\ga_5\fr{b_0 p_j}{m}$ is hermitian in Minkowski spacetime 
but nonhermitian in curved spacetime. 
However, 
the combination 
$\ga^j\ga_5 
 \left( \frac{-b_0 p_j+\frac{1}{2}i\prt_j b_0}{m}
 \right)$, 
which appears in the nonrelativistic hamiltonian, 
is hermitian in both Minkowski and curved spacetimes. 

All terms that appear in the Minkowski hamiltonian 
also appear in the curved-spacetime hamiltonian 
as part of a combination like this, 
confirming the hermiticity of the nonrelativistic hamiltonian. 

\subsection{Sensitivity of Completed Experiments}
We may exploit the correspondence between derivatives of SME coefficients 
in curved spacetime 
and effective values of coefficients in Minkowski coefficients 
to extract bounds on the former from published analysis of the latter. 
For example, 
the neutron-associated coefficient $\tb_X$ has the bound\cite{Brown:2010dt} 
$\abs{\tb_X}\lsim 10^{-33}$ GeV. 
In Minkowski spacetime, 
$\tb_X$ receives a contribution from $md_{XT}$. 
In curved spacetime, 
the effective value of $md_{XT}$ receives a contribution 
from the curl of $\vec{a}$: 
$(md_{XT})_{\text{eff}} \supset \frac{1}{2m}(\prt_Y a_Z - \prt_Z a_Y)$. 
Therefore, we find that the $Z$ component of the curl of $\vec{a}$ 
has the bound 
$\abs{ \prt_Y a_Z - \prt_Z a_Y } \lsim 10^{-33} $ GeV$^2$ 
for neutrons. 

Similar analysis yields the bounds from completed experiments 
that are summarized in Table \ref{boundtable}. 
Bounds expressed in parentheses require somewhat stronger assumptions 
than those expressed without parentheses. 
\begin{table} 
\tbl{Maximal sensitivity to derivatives of SME coefficients 
from already-completed experiments.} 
{
\begin{tabular}{c||cc|cc|cc} 
Weakly-curved-spacetime & \multicolumn{6}{c}{Sensitivity/GeV$^2$ and Reference} \\ 
\text{coefficient} & \multicolumn{2}{c|}{\text{Electron}} 
 & \multicolumn{2}{c|}{\text{Proton}} & \multicolumn{2}{c}{\text{Neutron}} \\ \hline 
$\prt_{[X} a_{T]}$   & $10^{-29}$   & \tablecite{Heckel:2008hw}   
 & $-$ &                                & $10^{-26}$   & \tablecite{Cane:2003wp} \\ 
$\prt_{[Y} a_{T]}$   & $10^{-29}$   & \tablecite{Heckel:2008hw}   
 & $-$ &                                & $10^{-26}$   & \tablecite{Cane:2003wp} \\ 
$\prt_{[Z} a_{T]}$   & $10^{-29}$   & \tablecite{Heckel:2008hw}   
 & $-$ &                                & $10^{-27}$   & \tablecite{Cane:2003wp} \\ \hline 
$\prt_{[Y} a_{Z]}$   & $10^{-34}$   & \tablecite{Heckel:2008hw}   
 & $10^{-33}$ & \tablecite{Kornack:2008zz}   & $10^{-33}$   & \tablecite{Brown:2010dt}   \\ 
$\prt_{[Z} a_{X]}$   & $10^{-34}$   & \tablecite{Heckel:2008hw}  
 & $10^{-33}$ & \tablecite{Kornack:2008zz}   & $10^{-33}$   & \tablecite{Brown:2010dt}   \\ 
$\prt_{[X} a_{Y]}$   & $10^{-32}$   & \tablecite{Heckel:2008hw}  
 & $10^{-28}$ & \tablecite{Peck:2012pt}      & $10^{-29}$   & \tablecite{Brown:2010dt}   \\ 
$\de^{JK}\prt_J a_K$ & $(10^{-21})$ & \tablecite{Altschul:2010na} 
 & $(10^{-11})$ & \tablecite{Kostelecky:2010ze} & $(10^{-11})$ & \tablecite{Kostelecky:2010ze} \\ \hline 
$\prt_X b_T$         & $(10^{-34})$ & \tablecite{Heckel:2008hw}  
 & $(10^{-33})$ & \tablecite{Kornack:2008zz} & $(10^{-33})$ & \tablecite{Brown:2010dt} \\ 
$\prt_Y b_T$         & $(10^{-34})$ & \tablecite{Heckel:2008hw}  
 & $(10^{-33})$ & \tablecite{Kornack:2008zz} & $(10^{-33})$ & \tablecite{Brown:2010dt} \\ 
$\prt_Z b_T$         & $(10^{-32})$ & \tablecite{Heckel:2008hw}  
 & $(10^{-28})$ & \tablecite{Peck:2012pt}    & $(10^{-29})$ & \tablecite{Peck:2012pt}  \\ \hline
$\prt_{(Y} b_{Z)}$   & $(10^{-29})$ & \tablecite{Heckel:2008hw}  
 & $-$  &                               & $(10^{-26})$ & \tablecite{Cane:2003wp} \\ 
$\prt_{(Y} b_{Z)}$   & $(10^{-29})$ & \tablecite{Heckel:2008hw}  
 & $-$  &                               & $(10^{-26})$ & \tablecite{Cane:2003wp} \\ 
$\prt_{(Z} b_{X)}$   & $(10^{-29})$ & \tablecite{Heckel:2008hw}  
 & $-$  &                               & $-$          \\ 
$\prt_{(Z} b_{X)}$   & $(10^{-29})$ & \tablecite{Heckel:2008hw}  
 & $-$  &                               & $(10^{-26})$ & \tablecite{Cane:2003wp} \\ 
$\prt_{(X} b_{Y)}$   & $(10^{-29})$ & \tablecite{Heckel:2008hw}  
 & $-$  &                               & $(10^{-27})$ & \tablecite{Cane:2003wp} \\ 
$\prt_{(X} b_{Y)}$   & $(10^{-29})$ & \tablecite{Heckel:2008hw}  
 & $-$  &                               & $(10^{-27})$ & \tablecite{Cane:2003wp} \\ \hline 
$\prt_{[Y} b_{Z]}$   & $10^{-21}$   & \tablecite{HohenseeAltschul} 
 & $10^{-20}$ & \tablecite{Wolf:2006uu}      & $10^{-5}$    & \tablecite{Kostelecky:2010ze} \\ 
$\prt_{[Y} b_{Z]}$   & $(10^{-29})$ & \tablecite{Heckel:2008hw}   
 & $-$  &                               & $(10^{-26})$ & \tablecite{Cane:2003wp} \\ 
$\prt_{[Z} b_{X]}$   & $10^{-21}$   & \tablecite{HohenseeAltschul} 
 & $10^{-20}$ & \tablecite{Wolf:2006uu}      & $10^{-5}$    & \tablecite{Kostelecky:2010ze}    \\ 
$\prt_{[Z} b_{X]}$   & $(10^{-29}$) & \tablecite{Heckel:2008hw}  
 & $-$  &                               & $(10^{-26})$ & \tablecite{Cane:2003wp} \\ 
$\prt_{[X} b_{Y]}$   & $10^{-23}$   & \tablecite{HohenseeAltschul} 
 & $10^{-20}$ & \tablecite{Wolf:2006uu}      & $10^{-5}$    & \tablecite{Kostelecky:2010ze}    \\ 
$\prt_{[X} b_{Y]}$   & $(10^{-29})$ & \tablecite{Heckel:2008hw}  
 & $-$  &                               & $(10^{-27})$ & \tablecite{Cane:2003wp} \\ 
\end{tabular} 
}
\label{boundtable} 
\end{table}

\section*{Acknowledgments}
This work was supported in part by Berry College 
and the Indiana University Center for Spacetime Symmetries (IUCSS).

\end{document}